\documentclass[a4paper,11pt]{article}
\usepackage{amsmath,fullpage}
\usepackage{epsfig}
\def\negcdot{\negmedspace\cdot\negthinspace}

\begin{document}

\begin{titlepage}
%\begin{flushright}

%corrected on June 13

%\end{flushright}

\vspace{1cm}
\begin{center}
{\Large \bf  The Search for New Physics in
$D^0 \to \gamma  \gamma  $  Decay}\footnote{Contributed Paper for
LPO1}\\

\vspace{1cm}
{\large \bf S. Fajfer$^{a,b}$,  P. Singer$^{c}$, J. Zupan$^a$}\\

\vspace{1cm}
{\it a) J. Stefan Institute, Jamova 39, P. O. Box 3000, 1001 Ljubljana,
Slovenia}
\vspace{.5cm}

{\it b)
Department of Physics, University of Ljubljana, Jadranska 19, 1000
Ljubljana,
Slovenia}
\vspace{.5cm}

{\it c) Department of Physics, Technion - Israel Institute  of
Technology,
Haifa 32000, Israel}

\end{center}

\vspace{0.25cm}

\centerline{\large \bf ABSTRACT}

\vspace{0.2cm}

We present main results of the investigation of the rare decay mode
$ D^0 \to \gamma  \gamma  $, in which the long distance contributions
are
expected to be dominant.
Using the Heavy Quark Chiral Lagrangian we have
considered the anomaly contribution which relates to the
annihilation part of the weak Lagrangian
 and  the one - loop $\pi$, $K$ diagrams. The loop contributions
 which are proportional to $g$ and contain
the $a_1$ Wilson coefficient are found to dominate the
decay amplitude.
The  branching ratio is then calculated to be
  $(1.0 \pm 0.5) \times 10^{-8}$.
Observation of an order of magnitude larger
branching ratio could signal new physics.

\end{titlepage}

%We present main results  of the analyses of the
% $D^0\to \gamma \gamma$ decay  within Standard Model.
%The details of the calculation are given in \cite{FSZ}.

In the past years the rare decays of B mesons came under the spotlights
as
 {\it the} source of possible signals of new physics.
 In the meanwhile
studies of rare D decays have received much less attention.
Partially this is because theoretical investigations of D weak decays
are
rather difficult, also due to the presence of many resonances close to
this
energy region. The penguin effects on the other hand, which are very
important in B and also in K decays, are  usually suppressed in the case
of charm mesons due to the presence of $d$, $s$, $b$ quarks in the loop
with the
respective values of CKM elements.

  Nevertheless, D meson physics has produced some interesting results in
the past year. Experimental results on time dependent decay rates of
$D^0 \to K^+ \pi^-$ by CLEO \cite{CLEO1} and $D^0 \to K^+ K^-$ and
$D^0 \to K^- \pi^+$ by FOCUS \cite{FOCUS1}
have stimulated several studies on the $D^0 - \bar D^0$ oscillations
\cite{Nir}. The recently measured D* decay width by CLEO \cite{CLEO2}
has provided the long
expected information on the value of $D^*D \pi$ coupling. Among the rare
D
decays, the decays $D\to V
\gamma$ and $D \to V(P) l^+ l^-$  are subjects of CLEO
and FERMILAB searches \cite{E791}. On the theoretical side, these rare
decays
of charm mesons into light vector meson and photon or lepton pair have
been considered lately by several authors (see, e.g.,
\cite{FPS}-
%,\cite{BGHP},\cite{HP},\cite{Khodj},\cite{Fajfer-98},%
\cite{Lebed-00},
for radiative leptonic D meson decay see \cite{Geng}). The
investigations of
$D\to V \gamma$ showed that certain branching ratios can be as large as
$10^{-5}$, like for $D^0\to \bar{K}^{*0} \gamma$, $D_s^+\to \rho^+
\gamma$ \cite{FPS,Lebed-00}.
However, the decays which are of some relevance to the $D^0\to 2 \gamma$
mode
studied here, like $D^0\to \rho^0 \gamma$, $D^0\to \omega \gamma$, are
expected with
branching ratios in the $10^{-6}$ range \cite{Fajfer-00}. Thus, it is
hard to believe that the
branching ratio of the $D^0\to 2 \gamma$ decay mode can be as high as
$10^{-5}$
in the Standard Model (SM), as found by \cite{RG}.
Apart from this
estimation, there
has been  no other detailed work on $D^0\to 2 \gamma$
 prior to our analysis \cite{FSZ}.
 In addition to these theoretical studies there are experimental
attempts to observe this decay rate done by CLEO and FOCUS \cite{Selen}.

 On the other hand, in the B and K meson systems there are numerous
studies
of their decays to two photons. For example, the $B_s\to \gamma \gamma$
decay
has been studied with various approaches within SM and beyond. In SM,
the
short distance (SD) contribution \cite{LSH} leads to a branching ratio
$B(B_s \to \gamma \gamma) \simeq
3.8 \times 10^{-7}$. The QCD corrections enhance this rate to $5 \times
10^{-7}$ \cite{RRS}.
On the other hand, in some of the SM extensions the branching ratio can
be
considerably larger. The two Higgs doublet scenario, for example, could
enhance this branching ratio by an order of magnitude \cite{BI}. Such
"new physics"
effects could at least in principle be dwarfed by long distance (LD)
effects.
However,  existing calculations show that these are not larger than the
SD
contribution \cite{ELLIS}, which is typical of the situation in
radiative B decays \cite{Eilam}. In the $K^0$ system the situation is
rather different. Here, the SD
contribution is too small to account for the observed rates of $K_S\to 2
\gamma$,
$K_L\to 2 \gamma$ by factors of $\sim 3-5$ \cite{Gaillard}, although it
could be of relevance in the
mechanism of CP-violation. Many detailed calculations of these processes
have
been performed over the years (see recent refs.
\cite{Gaillard}
%\cite{Goity}\cite{Amb}
-\cite{Kamb} and refs. therein),
especially using the chiral approach to account for the pole diagrams
and
the loops. These LD contributions lead to rates which are compatible
with
existing measurements.

Motivated by the experimental efforts to observe rare D meson
decays \cite{Selen}, and noticing that $B_s \to \gamma \gamma$ offers
possibility to
observe
physics beyond the SM, we undertook
an investigation of the $D^0\to  \gamma \gamma$ decay \cite{FSZ}.
Here we present only the main results of our analyses, while
the details of our work are presented in \cite{FSZ}.

The short
distance
contribution
is expected to be rather small, as already encountered in the one photon
decays \cite{FPS,BGHP}, hence the main contribution would come from long
distance
interactions. In order to treat the long distance contributions, we use
the
heavy quark effective theory combined with chiral perturbation theory
(HQ$\chi$PT)
\cite{wise}. This approach was used before for treating $D^*$ strong and
electromagnetic
decays \cite{itchpt}-\cite{GS}. The leptonic and semileptonic decays of
D meson were also treated
within the same framework (see \cite{itchpt} and references therein).

   The approach of HQ$\chi$PT introduces several coupling constants that
have
to be determined from experiment. The recent measurement of the $D^*$
decay
width \cite{CLEO2} has determined the $D^*D\pi$ coupling, which is
related to g, the basic
strong coupling of the Lagrangian. There is more ambiguity, however,
concerning
the value of the anomalous electromagnetic coupling, which is
responsible for
the $D^*D\gamma$ decays \cite{stewart,GS} (for further discussion on
this point see \cite{FSZ}).

   Let us address now some issues concerning the theoretical framework
 used in our treatment. For the weak vertex we  used the
factorization of
weak currents with nonfactorizable contributions coming from chiral
loops. The
typical energy of intermediate pseudoscalar mesons is of order $m_D/2$,
so that
the chiral expansion $p/\Lambda_\chi$ (for $\Lambda_\chi \simeq 1$
GeV)  is rather close to
unity. Thus, for the decay under study
we extend the possible range of
applicability of the chiral expansion of HQ$\chi$PT, compared to
previous treatments
like $D^*\to D \pi$, $D^*\to D \gamma$ \cite{stewart} or $D^*\to D
\gamma \gamma$ \cite{GS}, in which a heavy meson
appears in the final state, making the use of chiral perturbation theory
rather natural. The suitability of our undertaking here must be
confronted
with experiment, and possibly other theoretical approaches.

  At this point we also remark that the contribution of the order ${\cal
O}(p)$ does
not exist in the $D^0\to \gamma \gamma$ decay, and the amplitude starts
with
contribution of the order ${\cal O}(p^3)$. At this order the amplitude
receives an
annihilation type contribution proportional to the $a_2$ Wilson
coefficient, with the Wess-Zumino anomalous term coupling light
pseudoscalars
to two photons. However, the total amplitude is dominated by
terms
proportional to $a_1$ that contribute only through loops with Goldstone
bosons. Loop contributions proportional to $a_2$ vanish at this order.
We point out that any other model which does not involve intermediate
charged states cannot give this kind of contribution. Therefore, the
chiral
loops naturally include effects of intermediate meson exchange.

  The chiral loops of order ${\cal O}(p^3)$ are finite, as they are in
the similar
case of $K\to \gamma \gamma$ decays \cite{Gaillard}-\cite{Kamb}. The
next to leading terms might be
almost of the same order of magnitude compared to the leading ${\cal
O}(p^3)$ term,
the expected suppression being approximately $p^2/\Lambda^2_\chi$. The
inclusion of next order terms in the chiral expansion is not
straightforward
in the present approach. We include, however, terms which contain the
anomalous
electromagnetic coupling, and appear as next to leading order terms in
the
chiral expansion, in view of their potentially large contribution (as in
$B^*(D^*)\to B(D) \gamma \gamma$ decays considered in \cite{GS}). As it
turns out, these
terms are suppressed compared to the leading loop effects, which at
least
partially justifies the use of HQ$\chi$PT for the decay under
consideration.
Contributions of the same order could arise from light resonances like
$\rho$,
$K^*$, $a_0(980)$, $f_0(975)$. Such resonances are sometimes treated
with hidden
gauge symmetry (see, e.g., \cite{itchpt}), which is not compatible with
chiral perturbation
symmetry. Therefore, a consistent calculation of these terms is beyond
our
scheme and we disregard their possible effect.

The invariant amplitude for $D^0 \to \gamma \gamma $ decay can
 be written using gauge and Lorentz invariance in the following
 form:
\begin{equation}
M = \Big[ i M^{(-)} \big(g^{\mu \nu} -\frac{k_2^\mu k_1^\nu}{k_1
\negcdot k_2} \big)+
 M^{(+)} \epsilon^{\mu \nu\alpha\beta}k_{1\alpha}k_{2\beta}\Big]
 \epsilon_{1\mu}\epsilon_{2\nu},\label{eq-104}
\end{equation}
where $M^{(-)}$ is a parity violating  and $M^{(+)}$
a parity conserving  part of the amplitude,
while $k_{1(2)}$, $\epsilon_{1(2)}$ are respectively the
four momenta and the polarization vectors of the outgoing
photons.

\begin{figure}
\begin{center}
\epsfig{file=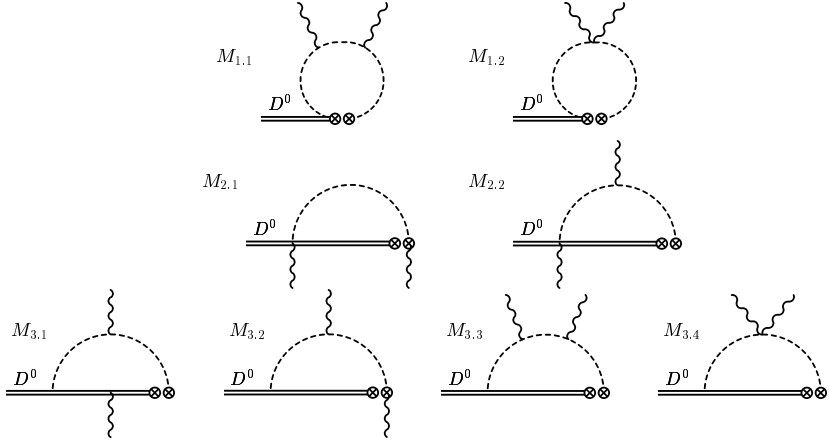, height=6cm}
\caption{One loop diagrams, not containing beta-like terms,
that give nonvanishing contributions to the
$D^0\to
\gamma \gamma$ decay amplitude. Each sum of the  amplitudes on
diagrams in one row $M_i=\sum_j
M_{i.j}$ is gauge invariant and finite. Numerical values are
listed in
Table \ref{tab-1}. }\label{fig-2}
\end{center}
\end{figure}

\begin{figure}
\begin{center}
\epsfig{file=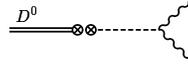, width=2.5cm}
\caption{Anomalous contributions to $D^0\to \gamma\gamma$ decay.
The intermediate pseudoscalar mesons propagating from the weak
vertex are $\pi^0,
\eta,\eta'$. }\label{fig-3}
\end{center}
\end{figure}

In the  discussion of weak radiative decays
$q' \to q \gamma \gamma$ or $q' \to q \gamma $ decays,
usually
the short (SD) and long distance (LD) contribution are separated.
The SD contribution in these transitions is a  result of the
penguin-like transition, while the long distance contribution
arises in
particular pseudoscalar meson decay as a result of the
nonleptonic four quark weak Lagrangian, when the photon is emitted
from the
quark legs. Here we follow this classification.
In the case of $b \to s \gamma \gamma$ decay
\cite{HK}  it was noticed that
without QCD corrections the rate
$\Gamma (b \to s \gamma \gamma)/ \Gamma (b \to s \gamma)$
is about $10^{-3}$. One expects that  a similar effect will show up
in the case of $c \to u \gamma \gamma$ decays.
Namely, according to the result of \cite{HK}
the largest contribution to $c \to u \gamma \gamma$ amplitude
would arise from the  photon emitted either from $c$ or $u$
quark legs
in the case of the  penguin-like  transition
$c \to u \gamma$.
Without QCD corrections the branching ratio for
$c \to u \gamma$  is rather suppressed,
being of the order $10^{-17}$ \cite{BGHP,HP}.
The QCD corrections \cite{GMW} enhance it up to order of $10^{-8}$.

 In our approach we include the $c \to u \gamma$ short distance
contribution by using the Lagrangian
 \begin{equation}
 {\cal L}=-\frac{G_f}{\sqrt{2}}V_{us} V_{cs}^*
C_{7\gamma}^{eff}\frac{e}{4\pi^2}
F_{\mu \nu}
  m_c \; \big(\bar u \sigma^{\mu\nu} \tfrac{1}{2}(1 + \gamma_5) c\big),
\label{eq-103}
 \end{equation}
 where $m_c$ is a charm quark mass.  In our analysis we follow
 \cite{GMW,Sasathesis}
 and we take $C_{7\gamma}^{eff}= (-0.7+ 2 i) \times 10^{-2}$.

 The main LD contribution will arise from the   effective four quark
nonleptonic $\Delta C = 1$
  weak Lagrangian
  %relevant for the $D^0\to \gamma\gamma$ decay
given by
\begin{equation}
{\cal L}=-\frac{G_f}{\sqrt{2}} \sum_{q=d,s}V_{uq}
 V_{cq}^* \big[ a_1  \big(\bar{q} \Gamma^\mu c) (\bar{u}
\Gamma_\mu q)+
 a_2 (\bar{u}\Gamma^\mu c) (\bar{q} \Gamma_\mu q)\big],
\label{eq-107}
\end{equation}
where $\Gamma^\mu=\gamma^\mu(1-\gamma_5)$, $a_i$ are effective
Wilson
coefficients \cite{buras}, and $V_{q_i q_j}$ are CKM matrix
elements. At
this point it is worth pointing out that long distance interactions will
contribute only if the $SU(3)$
flavor symmetry  is broken, i.e. if $m_s\neq m_d$. Namely, due to
$V_{ud}V_{cd}^* \simeq -V_{us}V_{cs}^*$ and $m_d = m_s$
the contribution arising from the weak Lagrangian (\ref{eq-107})
disappears in the case of exact $SU(3)$ flavor symmetry.

Going from quark to meson level effective Lagrangian one uses heavy
quark symmetry for c-quark and chiral symmetry of light quarks to
construct HQ$\chi$PT Lagrangian \cite{FSZ}. This is then used to calculate the
$D^0\to \gamma\gamma$ deacy width to one loop order. Leaving out the
details of  our calculation (see \cite{FSZ}), we discuss the
final results.

 The decay width for the $D^0\to \gamma \gamma$ decay can be obtained
 using the amplitude decomposition in \eqref{eq-104}
\begin{equation}
\Gamma_{D^0\to \gamma \gamma}= \frac{1}{16 \pi m_D} ( |M^{(-)}|^2+
\frac{1}{4} |M^{(+)}|^2 m_D^4) . \label{eq-106}
\end{equation}

\begin{figure}
\begin{center}
\epsfig{file=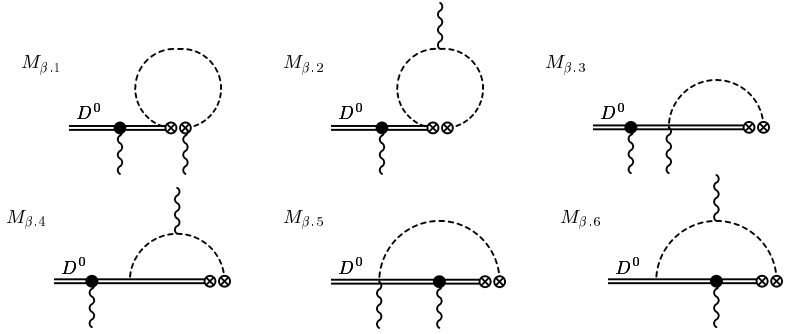, height=4.5cm}
\caption{The diagrams which give nonzero amplitudes
with one $\beta$-like coupling (denoted by $\bullet$). }\label{fig-5}
\end{center}
\end{figure}

The main contribution to the
decay width arises from the diagrams presented on
 Figs. \ref{fig-2}, \ref{fig-3}.  The calculated amplitudes
 depend on
the number of input
 parameters, as mentioned
in \cite{FSZ}.
The coupling $g$ is extracted from existing
experimental data  on $D^*\to D\pi$.  Recently CLEO Collaboration has
obtained the
first measurement of $D^{*+}$ decay width $\Gamma(D^{*+})=96\pm 4\pm 22$
$\rm{ keV}$ \cite{CLEO2} by studying the $D^{*+}\to D^0 \pi^+$. Using the
value of
decay width together with
branching ratio $Br(D^{*+}\to D^0 \pi^+)=(67.7\pm 0.5)\% $   one
immediately finds at tree level that $g=0.59 \pm 0.08 $.
The chiral corrections to this coupling were found to contribute about
$10\%$  \cite{itchpt,stewart}.  In order
to obtain the $\alpha$ coupling, we use present experimental
data on
$D_s$ leptonic decays ($f_{D}\simeq f_{D_{s}}=\alpha/\sqrt{m_D}$).   In our
calculation we take $\alpha=0.31\; {\rm GeV}^{3/2}$ \cite{FSZ}.
For the Wilson coefficients $a_1$ we take $1.26$
%of(\ref{Lquark})
and $a_2=-0.47$ \cite{buras}.
We give here the numerical results for the
one-loop amplitudes in Table 1.

\begin{table} [h]
\begin{center}
\begin{tabular}{|l|r l|r l|} \hline
 &$M_{ i}^{(-)}$&$ [\times 10 ^{-10}{\rm  \;GeV}]$& $M_{i}^{(+)}$&$
[\times 10^{-10}{\rm  \;GeV^{-1}}]$\\ \hline\hline
Anom. & $0$ & & $-0.53$& \\ \hline
SD & $-0.27$&$-0.81 i$ & $-0.16$&$ -0.47 i$ \\ \hline
$1$ & $3.55$&$+9.36i$ & $0$&\\ \hline
$2$ &$1.67$ & &  $0$&  \\ \hline
$3$ & $-0.54$&$+2.84i$&$0$& \\ \hline
\hline
$\sum_i  M_i^{(\pm)}$& $4.41$&$+11.39 i$ &$-0.69$&$ -0.47 i$\\ \hline

\end{tabular}
\caption{\footnotesize{Table of the nonvanishing finite
amplitudes. The amplitudes coming from the
anomalous and short distance
($C_{7\gamma}^{eff}$) Lagrangians are presented. The finite
and  gauge invariant sums of
one-loop amplitudes are listed in the next three lines
($M_i^{(\pm)}=\sum_j M_{i.j}^{(\pm)}$).
The numbers $1,2,3$ denote the row of diagrams on the
Fig.\ref{fig-2}.
In the last line the sum of all amplitudes is given.}}
\label{tab-1}
\end{center}
\end{table}

In the determination of $D^*\to D \gamma\gamma $ and
$B^* \to B \gamma \gamma$ a sizable contribution from
$\beta$-like electromagnetic terms \cite{FSZ} % \eqref{eq-100}
has been
found \cite{GS}. Therefore, we have to investigate their
effect
in the $D^0 \to \gamma \gamma $ decay amplitude.
The nonzero parity violating  parts of the one loop
diagrams containing
$\beta$
coupling are given on Fig. \ref{fig-5}, while numerical results are
presented in Table 2.

\begin{table} [h]
\begin{center}
\begin{tabular}{|l|c|c|c|} \hline
Diag. &$M_{i}^{(-)} [\times 10 ^{-10}{\rm  \;GeV}]$& $M_{i}^{(+)}
[\times 10
^{-10}{\rm  \;GeV^{-1}}]$\\ \hline\hline
$\beta .1$ & $0$ & $-2.69$ \\ \hline
$\beta .2$ & $0$ & $2.69$\\ \hline
$\beta .3$ &$0$ &  $2.11$ \\ \hline
$\beta .4$ & $0.88$ &$-0.007$ \\ \hline
$\beta .5$ &$0$ &  $0.51$ \\ \hline
$\beta .6$ &$-2.88 $&$-0.52$ \\ \hline
\hline
$\sum_i  M_i^{(\pm)}$& $-2.00$ &$2.09$\\ \hline

\end{tabular}
\caption{\footnotesize{Table of  nonzero contributions of the
amplitudes coming from the diagrams
 with $\beta$ coupling. % (Fig. \ref{fig-5}).
 % to the corresponding one chiral loop amplitudes.
In the last line the sums of the contributions are  presented.
We use $\beta=2.3 $ GeV${}^{-1}$, $m_c=1.4$ GeV. }}
\label{tab-2}
\end{center}
\end{table}
Using short distance contributions, the finite one loop diagrams and the
anomaly parts of the
amplitudes
and with numerical values of the amplitudes as listed in Table
\ref{tab-1},
one obtains
\begin{equation}
Br(D^0\to \gamma\gamma)=1.0 \times 10^{-8}.
\end{equation}
This result is slightly changed when one takes into account the terms
dependent on $\beta$. %\eqref{eq-100}.
The branching ratio obtained when
we
sum all contributions is
\begin{equation}
 Br(D^0\to \gamma\gamma)=0.95 \times 10^{-8}.
\end{equation}

By varying $\beta$ within $1 \;{\rm GeV}^{-1}
\le \beta \le 5 \; {\rm GeV}^{-1}$ and keeping
$g=0.59\pm 0.08$, the branching ratio
is changed by at most 10\%. On the other hand, one has to keep in mind
that
the loop contributions
involving $\beta$ are not finite and have to be regulated. We have used
$\overline{\rm MS}$ scheme, with  the
 divergent parts being absorbed by counterterms.
 The size of these is not
known, so they might influence the error in our
 prediction of the branching ratio.
Note also  that changing $\alpha$ would
affect the predicted
branching ratio.  For instance, if the chiral corrections do
decrease the value of $\alpha$  by $30\%$ this would decrease
 the predicted
branching ratio down to $0.5\times 10^{-8}$.

We have presented here the results of the  detailed calculation of the
decay
amplitude
$D^0\to \gamma \gamma$, which
includes short distance and long distance contributions, by making use
of
the theoretical
tool of Heavy Quark Chiral Perturbation Theory Lagrangian.
Within this
framework, the leading
contributions are found
to arise from the charged $\pi$ and $K$ mesons running in the chiral
loops, and
are  of order ${\cal O}(p^3)$.
These terms are finite and contribute only to the parity violating part
of
the amplitude. The
inclusion of terms of higher order in the chiral expansion is
unfortunately
 plagued
 with the uncertainty caused
by the lack of knowledge of the counterterms. As to the parity
conserving
part of the decay, it is given
by terms coming from the short distance contribution, the anomaly and
from
loop terms containing
the beta coupling, the latter giving most of the amplitude. The size of
this
part of the amplitude
is approximately one order of magnitude smaller than the parity
violating
amplitude, thus
contributing less than 20\% to the decay rate. Therefore, our
calculation
predicts that the $D\to 2 \gamma$
decay is mostly a parity violating transition.

 In addition to the uncertainties we have mentioned, there is the
question
of the suitability of
the chiral expansion for the energy involved in this process; the size
of
the uncertainty related
to this is difficult to estimate. Altogether, our estimate is that the
total
uncertainty is not larger
than 50\%. Accordingly, we conclude that the predicted branching ratio
is
\begin{equation}
Br(D^0\to \gamma \gamma)= (1.0 \pm 0.5)\times 10^{-8}. \label{fin-res}
\end{equation}
 The reasonability of this result can be deduced also from a comparison
with the calculated decay
rates for the $D^0\to \rho(\omega)\gamma$, which are found to be
expected
 with
a branching ratio of
approximately $10^{-6}$ \cite{FPS,BGHP,Fajfer-00}.

 We look forward to experimental attempts of detecting this decay. Our
result suggests that the
observation of $D\to 2 \gamma$ at a rate which is an order of magnitude
larger
than \eqref{fin-res}, could be a
signal for the type of "new physics", which leads to
sizable enhancement \cite{Sasathesis}
of the short-distance $c \to u \gamma$ transition.

\end{document}